\documentstyle[12pt]{article}
\def \be{\begin{equation}}
\def \ee{\end{equation}}
\def \lb{\label}
\def \ba{\begin{array}{l}}
\def \ea{\end{array}}
\def \ov{\over}
\def \s{\sigma}
\def \t{\tau}
\def \a{\alpha}
\def \b{\beta}
\def \d{\delta}
\def \D{\Delta}

\def \f{\phi}

\def \G{\Gamma}
\def \lm{\lambda}
\def \n{\nabla}
\def \p{\varphi}
\def \tl{\tilde}
\def \ol{\overline}

\def \la{\langle}
\def \ra{\rangle}

\def \fr{\frac}
\def \R{R_{c}}
\def \T{T_{c}}

\def \2{\frac{1}{2}}
\def \4{\frac{1}{4}}

\newcommand{\bi}{\bibitem}
\newcommand{\pub}[4]{{\em #1 }{\bf #2}, #3 (#4)}

\newcommand{\jpa}{J. Phys. A}

\def \(({\left(}
\def \)){\right)}
\def \[[{\left[}
\def \]]{\right]}

\def \bi{\bibitem}

\newcommand{\sectio}[1]{\section{#1}\setcounter{equation}{0}}

\begin{document}

\input epsf
\title{\bf
 Vector breaking of replica symmetry in some low temperature disordered
systems}
\vskip 3 true cm

\author{Viktor Dotsenko$^{(1)}$ and Marc M\'ezard}

\date{\it
Laboratoire de Physique Th\'eorique de l'Ecole Normale Sup\'erieure
 \footnote {Unit\'e propre du CNRS,  associ\'ee
 \`a\ l'Ecole
 Normale Sup\'erieure et \`a\ l'Universit\'e de Paris Sud} , \\
24 rue
 Lhomond, 75231 Paris Cedex 05, France\\
(1) Permanent address: Landau Institute for Theoretical Physics, \\
2 Kosygina str., 117940 Moscow, Russia}

%\date{\today}

\maketitle

\begin{abstract}
We present a new method to study disordered systems in the low temperature
limit.
The method uses the replicated Hamiltonian. It studies the saddle points
of this Hamiltonian and  shows how the various saddle point contributions
can be resummed in order to obtain the scaling behaviour at low temperatures.
In a large class of strongly disordered systems, it is necessary to include
saddle points of the Hamiltonian which break the replica symmetry in a vector
sector,
as opposed to the usual matrix sector breaking of spin glass mean field theory.

LPTENS preprint 96-62
\end{abstract}

\sectio{Introduction}

The use of the replica method has turned out to be very efficient in some
disordered systems. It allows for a detailed characterization of
the low temperature phase at least at the mean field level. In all the
mean field spin glass like problems where one can expect  the
mean field theory to be exact, the Parisi scheme of replica symmetry
breaking \cite{mpv} is successful, and at the moment there is no
counterexample showing that it does not work. On the other hand, the
low temperature phase of these systems is complicated enough, even
at the mean field level. One might hope that the very
low temperature limit
could be easier to analyse, while its physical content should be
basically the same. This very low temperature limit is also an extreme case
where one
might hope to get a better understanding of the finite dimensional problem.
At first sight the low temperature limit is indeed simpler since
the partition function could be analysed at the level of
a saddle point approximation. However it is easy to see that generically
this limit does not commute with the limit of the number of
replicas going to zero. There is a very basic origin to this
non commutation, namely the fact that  there still exist, even at zero
temperature,
sample to sample fluctuations. In this paper we try to develop
a method of summation over all saddle point in replica space,
in order to get the low temperature behaviour of glassy systems.
The main
aim of this paper is  to propose this new method. We have
tested it on some elementary problems which can be solved directly.
As for its application to more difficult problems, we have also
obtained some very good approximation to the zero temperature
fluctuations  a
particle in a random medium, as well as some interesting
scaling relations in the random field Ising model.
Sect.2 presents the method and illustrates it on a variety of
zero dimensional problems. In section 3 we discuss the case of directed
polymers in random media with long range interactions, where we rederive
the scaling exponents
using this new method. In sect.4 we study the D dimensional random field Ising
model.
Perspectives are briefly summarized in sect.5.

\sectio{Zero-dimensional systems}

\subsection{The Ising Model}

To demonstrate in the  simplest terms how the proposed procedure works,
we consider first some trivial  zero dimensional problems.
The simplest example is  one Ising spin $\s = \pm 1$ in a random field $h$.
The Hamiltonian is:

\be
\lb{aaa}
H = \s h
\ee
where the distribution for the random field is Gaussian:

\be
\lb{aab}
P(h) = \fr{1}{\sqrt{2\pi h_{0}^{2}}} \exp(-\fr{h^{2}}{2h_{0}^{2}})
\ee
The  free energy  is:

\be
\lb{aac}
-\b F(h_{0};\b) =
\ol{\ln\left[\sum_{\s=\pm 1}\exp(-\b \s h)\right] } =
\int_{-\infty}^{+\infty}Dx
\ln[1 + \exp(2 \b h_{0} x)]
\ee
where Dx is the centered gaussian measure of width one:
$Dx = \fr{dx}{\sqrt{2\pi}} \exp(-\2 x^{2})$.
In particular, in the zero temperature limit one  finds:

\be
\lb{aad}
F(h_{0};\b\to\infty) = \fr{2 h_{0}}{\sqrt{2\pi}}
\ee

Let us consider now how this "problem" can be solved in terms of the replica
approach:

\be
\lb{aae}
\ba
-\b F(h_{0};\b) = \lim_{n\to 0}\fr{1}{ n}(\ol{Z^{n}} - 1) =\\
\\
\lim_{n\to 0}\fr{1}{ n}
\left[ \sum_{\{\s_{a}\}=\pm 1}
\exp\{\2 \b^{2}h_{0}^{2}(\sum_{a=1}^{n}\s_{a})^{2}\} - 1 \right] =\\
\\
lim_{n\to 0}\fr{1}{ n} \left[ \sum_{k=0}^{n}\fr{n!}{k!(n-k)!}
\exp\{\2 \b^{2}h_{0}^{2}(2k-n)^{2}\} - 1 \right]
\ea
\ee
In view of the application of the method to more complicated problems
we want to compute the behaviour at low temperature. This cannot be done
naively from
a saddle point evaluation of the sum at large $\b$, because of the
non commutativity of the limits $\b \to \infty$ and $n \to 0$. Instead we
proceed
as follows. The term $k=0$, which is the contribution from the
'replica symmetric (RS) configuration'
$\s_a=+1$, is singled out; its
 contribution is equal to $1 + O(n^{2})$, which
cancels the $(-1)$ in eq.(\ref{aae}). The contributions of the rest of the
terms (which could be interpreted as corresponding to the states
with  "replica symmetry breaking" (RSB) in the replica vector
$\{\s_{a}\}$) can be represented as follows:

\be
\lb{aaf}
F(h_{0};\b) = -\lim_{n\to 0}\fr{1}{\b n}
\sum_{k=1}^{\infty}\fr{\G(n+1)}{\G(k+1)\G(n-k+1)}
\exp\{\2 \b^{2}h_{0}^{2}(2k-n)^{2}\}
\ee
Here the summation over $k$ can be extended beyond $k=n$ to $\infty$ since the
gamma function is equal to infinity at negative integers.

Now we  perform the analytic continuation $n \to 0$,
Using the relation:

\be
\fr{\G(n+1)}{\G(k+1)\G(n-k+1)}|_{n \to 0} \simeq n {(-1)^{k-1} \over k}
\lb{aag}
\ee
Thus, for the free energy (\ref{aaf}) one obtains:

\be
\lb{aak}
\ba
- \b F(h_{0};\b) =
\sum_{k=1}^{\infty}\fr{(-1)^{k-1}}{k}
\exp( \b^{2}h_{0}^{2} k^{2} ) =
\int_{-\infty}^{+\infty}Dx
\ln\left[1 + \exp(2 \b h_{0}x) \right]
\ea
\ee
We see that this result coincides with the one (\ref{aac}) obtained by
the direct calculation. This is of course no surprise since we have just done
an exact replica computation. But it exemplifies some of the steps that we
shall
need below, in particular the proper definition and computation of
 the divergent series appearing
in (\ref{aak}) through an integral representation.

\subsection{The "Soft" Ising Model}

Consider now the "soft" version of the Ising model described by the
double-well Hamiltonian:

\be
\lb{aba}
H = -\2\t\f^{2} + \4\f^{4} - h\f
\ee
where the random field is described by the Gaussian distribution (\ref{aab}).
We concentrate again on the zero temperature limit. Besides, we  assume that
the typical value of the field $h_0$ is small ($h_0 \ll \t^{3/2}$).
In this case the field will not destroy the double-well structure
of the Hamiltonian (\ref{aba}), and (at $T\to 0$) the system must be
equivalent to the discrete Ising model considered before. (The "opposite
limit" of the random field Hamiltonian with only one ground state
 will be considered in Sec.2.3).

The direct calculation of the zero-temperature free energy is
trivial. For a given value of the field $h \ll \t^{3/2}$ the ground states
of the Hamiltonian (\ref{aba}) are:
$\f_1 \simeq +\sqrt{\t} + h/2\t$, for $h > 0$; and
$\f_1 \simeq -\sqrt{\t} + h/2\t$, for $h < 0$. In both cases the
corresponding energy is $E_g(h) \simeq -\4 \t^2 - |h|\sqrt\t$.
Thus,  the zero-temperature averaged free energy is:

\be
\lb{abb}
F(h_0) \simeq -\4 \t^{2} - 2 \sqrt{\t}
\int_{0}^{+\infty} \fr{dh}{\sqrt{2\pi h^{2}_{0}}} h \exp (-\fr{h^{2}}{2h_{0}})
=
-\4 \t^{2} -\fr{2 h_{0} \sqrt{\t}}{\sqrt{2\pi}}
\ee

Consider now how this result can be obtained in terms of replicas.
The replica Hamiltonian and the corresponding saddle-point equations are:

\be
\lb{abc}
H_n =  -\2\t\sum_{a=1}^{n}\f_{a}^{2} + \4\sum_{a=1}^{n}\f_{a}^{4}
-\2\b h_{0}^{2} (\sum_{a=1}^{n} \f_{a})^{2}
\ee
\be
\lb{abd}
-\t\f_{a} + \f_{a}^{3} = \b h_{0}^{2} (\sum_{a=1}^{n} \f_{a})
\ee

The "replica-symmetric" solution of these equations (in the limit $n\to 0$)
is: $\f_{a} = \f_{rs} = \sqrt{\t}$. The corresponding energy is
$E_{rs} = -\4 n \t^{2}$. This solution (in the limit $n\to 0$)
does not involve the contribution from the random field.

Proceeding along the lines of the Section 2.1, we have to look also
for the solution of the eqs.(\ref{abd}) which would involve the
"replica symmetry breaking" in the replica vector $\{\f_{a}\}$:

\be
\lb{abe}
\f_{a} = \left\{ \begin{array}{ll}
                 \f_{1} & \mbox{for $a = 1, ..., k$}\\
                          \f_{2} & \mbox{for $a = k+1, ..., n$}
                          \end{array}
                 \right.
\ee
In terms of this Ansatz in the limit $n\to 0$ the replica summations
can be performed according to the following simple rule:
$\sum_{a=1}^{n} \f_{a} = k \f_{1} + (n-k)\f_{2} \to k(\f_{1} - \f_{2})$.
The saddle-point eqs.(\ref{abd}) then turn into two equations for
$\f_{1}$ and $\f_{2}$:

\be
\lb{abdd}
-\t\f_{1,2} + \f_{1,2}^{3} = \b k h_{0}^{2} (\f_{1} - \f_{2})
\ee
Assuming that $\b k h_{0}^{2} \ll \t$
(the explanation of this strange assumption  -considering that
we are interested in the $\b \to \infty$ limit!- will be given below),
in the leading order one gets:

\be
\lb{abf}
\f_{1} \simeq + \sqrt{\t} \; ; \; \; \;  \f_{2} \simeq - \sqrt{\t}
\ee
 From the eq.(\ref{abc}) one obtains the corresponding energy of the
above "RSB" saddle-point solution:

\be
\lb{abg}
\ba
E_{k} = -\2 k \t (\f_{1}^{2} - \f_{2}^{2}) + \4 k (\f_{1}^{4} - \f_{2}^{4})
-\2\b h_{0}^{2} k^{2} (\f_{1} - \f_{2})^{2} \simeq \\
\\
\simeq - 2 \b k^{2} h_{0}^{2} \t \; \; + \; O(h_{0}^{4})
\ea
\ee

Now, similarly to the calculations of Sec.2.1 for the zero-temperature
free energy one obtains, summing the contributions from all these
saddle points:

\be
\lb{abh}
\ba
F(h_{0}) = -\lim_{n\to 0}\fr{1}{\b n}(Z_{n} -1) \simeq \\
\\
-\lim_{n\to 0}\fr{1}{\b n}(Z_{"rs"} -1)
-\lim_{n\to 0}\fr{1}{\b n} Z_{"rsb"} = \\
\\
-\lim_{n\to 0}\fr{1}{\b n} [\exp(\4\b n \t^{2}) - 1] -
\fr{1}{\b} \sum_{k=1}^{\infty} \fr{(-1)^{k-1}}{k}
\exp(2 \b^{2} k^{2} h_{0}^{2} \t) =\\
\\
-\4\t^{2} - \fr{1}{\b} \int_{-\infty}^{+\infty} \fr{dx}{\sqrt{2\pi}}
\exp(-\2 x^{2}) \ln[1 + \exp(2\b h_{0} x\sqrt{\t})]
\ea
\ee
Taking the limit $\b \to \infty$ one finally gets the result:

\be
\lb{abi}
F(h_{0}) \simeq -\4\t^{2} -\fr{1}{\b} 2 \b h_{0}\sqrt{\t}
\int_{0}^{+\infty} \fr{dx}{\sqrt{2\pi}} x \exp(-\2 x^{2}) =
-\4\t^{2} -\fr{2 h_{0}\sqrt{\t}}{\sqrt{2\pi}}
\ee
which coincides with eq.(\ref{abb}).

It is worth to note that the summation of the series in eq.(\ref{abh})
can also be performed in the other way:

\be
\lb{abj}
F_{rsb} = -\fr{1}{\b} \sum_{k=1}^{\infty} \fr{(-1)^{k-1}}{k}
\exp(2 \b^{2} k^{2} h_{0}^{2} \t) =
\fr{1}{2i\b} \int_{C} \fr{dz}{z\sin(\pi z)}
\exp(2 \b^{2} z^{2} h_{0}^{2} \t)
\ee
where the integration goes over the  contour in the complex plane
shown in Fig.1a. Then we can move the  contour to the position shown
in Fig.1b, and after the change of integration variable:

\be
\lb{abk}
z \to [2 \b^{2} h_{0}^{2} \t]^{-1/2} i x
\ee
in the limit $\b \to \infty$ we have:

\be
\lb{abl}
\sin(\pi z) \simeq \fr{1}{\b} i \pi [2 h_{0}^{2} \t]^{-1/2} x
\ee
Then, taking into account also the contribution from the pole at $x=0$
for the integral in eq.(\ref{abj}) we get:

\be
\lb{abm}
F_{rsb} = \fr{h_{0}\sqrt{2\t}}{2\pi}
\int_{-\infty}^{+\infty} \fr{dx}{x^{2}} [\exp(-x^{2}) - 1] =
- \fr{2 h_{0} \sqrt{\t}}{\sqrt{2\pi}}
\ee
which again  coincides with the results (\ref{abb}) and (\ref{abi}).

This little exercise with the integral representation of the divergent
series in eq.(\ref{abj}) shows in particular that the "effective"
value of the parameter $\b k \to \b z$ which enter into the saddle-point
equations (\ref{abdd}) scales (according to (\ref{abk})) as
$(h_{0}\sqrt{\t})^{-1}$. That is why in the zero temperature limit
the "effective" value of the
factor $\b k h_{0}^{2} \sim h_{0}/\sqrt{\t}$ in the eq.(\ref{abdd})
can be assumed to be small compared to $\t$ (for small fields
$h_{0} \ll \t^{3/2}$).

\vspace{5mm}

{\it Replica fluctuations}

\vspace{5mm}

Because of the non commutativity of the limits $n \to 0$ and
$\b \to \infty$, one cannot get the exact result
by keeping only the saddle-point states of the replica Hamiltonian.
Actually, averaging over quenched disorder involves the effects of
sample to sample fluctuations which in terms of the replica formalism
manifest themselves as the contribution from the replica fluctuations.
In other words, to get exact result in terms of replicas the
contribution from the  saddle-points is not enough,
and  one has to integrate over replica fluctuations even in the
zero-temperature limit.

This phenomenon can be easily demonstrated for the above example
of the "soft" Ising model. Les us take into account the contribution
from the Gaussian replica fluctuations near the "replica-symmetric"
saddle-point $\f_{a} = \f_{rs} = \sqrt{\t}$:

\be
\lb{abn}
\f_{a} = \f_{rs} + \p_{a}
\ee
 From the eq.(\ref{abc}) for the "replica-symmetric" part of the partition
function we get:

\be
\lb{abo}
\ba
Z_{"rs"} = \exp(\4\b n \t^{2})
\int d\p_{a}
\exp\{ -\b \sum_{a,b}^{n}(\t\d_{ab} - \2\b h_{0}^{2})\p_{a}\p_{b} \}
\\
\simeq \exp\{ \4\b n \t^{2} + \fr{\b n h_{0}^{2}}{4\t} -
\2 n \ln(\b\t) \}
\ea
\ee
Therefore, in the zero-temperature limit one obtains the following
contribution to the free energy:

\be
\lb{abr}
F_{"rs"} = -\4 \t^{2} - \fr{h_{0}^{2}}{4\t}
\ee
We see that at $T = 0$ there exists a finite contribution $\sim h_{0}^{2}/\t$
due to the replica fluctuations.
In the particular example considered the value of $h_{0}$ was assumed
to be small, and this contribution can be treated as a small
correction. However, we should keep in mind that the
contribution from the replica fluctuations in general  could appear to be
of the same order as that from the saddle points. Therefore, the calculations
we are going to perform in next sections for less trivial examples taking
into account only saddle-point states can not pretend to give exact
results, giving only the scaling dependence from the parameters of a model.

\vspace{5mm}

{\it Saddle points}

\vspace{5mm}

In the above calculations of the free energy for the "soft" Ising
system we have taken into account only the contribution from the two
{\it minima} of the double-well potential. The existence of the
third saddle-point, which is the maximum at $\f=0$, has been ignored.
In this particular example such an algorithm looks natural. However,
in less trivial systems very often it is not easy to distinguish
the types of the saddle points involved. Moreover, it could be
very hard to impose a simple and robust "discrimination" rule with respect
to different types of  saddle-points, which would not block the
calculations at the very start.

Because of this, we would like to propose a somewhat modified scheme of
calculations
which takes into account {\it all} saddle points. In the above
example of the "soft" Ising model the third saddle-point (the maximum)
is at $\f = 0$. Then, instead of the Ansatz (\ref{abe}) let us represent
the replica vector $\f_{a}$ as follows:

\be
\lb{abs}
\f_{a} = \left\{ \begin{array}{ll}
                 +\sqrt{\t} & \mbox{for $a = 1, ..., k$}\\
                          -\sqrt{\t} & \mbox{for $a = k+1, ..., k+l$}\\
                 0 & \mbox{for $a = k+l+1, ..., n$}
                          \end{array}
                 \right.
\ee
For the corresponding "energy" (in the limit $n \to 0$) from the replica
Hamiltonian (\ref{abc}) one easily finds:

\be
\lb{abt}
H_{kl} = -\4 \t^{2} (k+l) - \2 \b h_{0}^{2} \t (k-l)^{2} \; \;
+ \; O(h^{4}_{0})
\ee
Note that in terms of the Ansatz (\ref{abs}) the "replica symmetric"
state ($k = l = 0$), $\f_{a} = \f_{0} = 0$ has  zero energy, so that it
gives no contribution to the free energy.

The
combinatoric factor in the $n \to 0$ limit is now:

\be
\lb{abu}
\fr{n!}{k! \; l! (n-k-l)!} \to n \fr{(-1)^{k+l-1}}{k+l} \fr{(k+l)!}{k! \; l!}
\ee
Thus, for the free energy (for $\b \to \infty$) we obtain:

\be
\lb{abv}
F(h_{0}) = -\fr{1}{\b} \sum_{k+l=1}^{\infty}
\fr{(-1)^{k+l-1}}{k+l} \fr{(k+l)!}{k! l!}
\exp\{ \4\b\t^{2}(k+l) + \2 \b^{2} h_{0}^{2} \t (k-l)^{2} \}
\ee
This series can be summed up in a similar way as the ones in
eqs.(\ref{aak}) and (\ref{abh}):

\be
\lb{abvv}
F(h_{0}) =
-\fr{1}{\b} \int_{-\infty}^{+\infty} Dx
\ln\left[ 1 + \exp\{\fr{\b}{4}\t^{2} + \b h_{0}\sqrt{\t} x \} +
\exp\{\fr{\b}{4}\t^{2} - \b h_{0}\sqrt{\t} x\} \right]
\ee
In the limit $\b\to\infty$ one finds:

\be
\lb{abvvv}
F(h_{0}) =
-\fr{1}{\b} \int_{-\infty}^{+\infty} Dx
\left[ \fr{\b}{4}\t^{2} + \b h_{0}\sqrt{\t} |x| \right] =
-\4 \t^{2} - \fr{2 h_{0} \sqrt{\t}}{\sqrt{2\pi}}
\ee
Again, we get the correct result. While it might seem at first sight
somewhat
 "magic", at least some aspects of this computation can be understood.
In the example considered
(as well as in the further examples to be studied below) the relevant states,
which contribute to the free energy, have negative energy $-E(h)$. Then,
in the low temperature
limit the partition function of a given sample is $Z \simeq \exp(+\b E(h))$.
Therefore, in the limit $\b \to \infty$ the free energy  can be represented
with  exponential
accuracy as follows:

\be
\lb{abw}
\ba
F(h_{0}) = -\fr{1}{\b} \ol{\ln Z} \simeq -\fr{1}{\b} \ol{\ln (1 + Z)}\\
\\
= -\fr{1}{\b} \sum_{m=1}^{\infty} \fr{(-1)^{m-1}}{m} \ol{Z^{m}}
\ea
\ee
One can easily check that after averaging $\ol{Z^{m}} \equiv Z_{m}$
and taking into account the contributions of the two minima of the
corresponding replica Hamiltonian $H_{m}$ one recovers the series
in eq.(\ref{abv}).

The only "magic" rule which should be followed in the direct replica
calculations is that the "background" state,
$\f_{0} = 0$ (the one with zero energy) in the Ansatz
for the replica vector $\f_{a}$ of the type
(\ref{abs}) should be placed in the {\it last} group of replicas.
Using this rule, the series obtained for the free energy will correspond
to the above interpretation (\ref{abw}).

\subsection{One-Well Potential}

Consider now how the method works in the case where the Hamiltonian
has only one minimum:

\be
\lb{abba}
H = \fr{1}{\a} \f^{\a} - h \f
\ee
where $\f \geq 0$ and $\a \geq 2$, and the random field $h$ is again described
by the Gaussian distribution (\ref{aab}). For $\a = 4$ this system can be
interpreted as the variant of the Hamiltonian (\ref{aba}) in the limit
of strong magnetic fields.

In the zero-temperature limit
the free energy is defined by the ground state $\f(h) = h^{1/(\a - 1)}$
for $h > 0$, and $\f = 0$ for $h \leq 0$.
Its energy is $E(h) = -\fr{\a - 1}{\a} h^{\a/(\a - 1)}$ for $h > 0$, and
$E = 0$ for $h \leq 0$. Therefore, for the averaged zero-temperature free
energy we find:

\be
\lb{abbb}
\ba
F(h_{0}) = -\fr{\a - 1}{\a}
\int_{0}^{+\infty} \fr{dh}{\sqrt{2\pi h_{0}^{2}}} h^ {\fr{\a}{\a - 1}}
\exp\{ -\fr{h^{2}}{2h_{0}^{2}} \} = \\
\\
-h_{0}^ {\fr{\a}{\a - 1}} \times \fr{\a - 1}{\a}
\int_{0}^{+\infty} \fr{dx}{\sqrt{2\pi}} x^ {\fr{\a}{\a - 1}}
\exp( -\2 x^{2} ) \equiv  A(\a) \times h_{0}^ {\fr{\a}{\a - 1}}
\ea
\ee
In terms of replicas, the corresponding replicated Hamiltonian is:

\be
\lb{abbc}
H_{n} = \fr{1}{\a} \sum_{a=1}^{n} \f_{a}^{\a} -
\2 \b h_{0}^{2} \sum_{a,b=1}^{n} \f_{a} \f_{b}
\ee
This Hamiltonian has a trivial "background" extremum at $\f = 0$ with
 zero energy. Therefore, following the scheme proposed in the
previous subsection, we  look for non-trivial saddle-point
solutions in terms of the following Ansatz:

\be
\lb{abbe}
\f_{a} = \left\{ \begin{array}{ll}
                 \f & \mbox{for $a = 1, ..., k$}\\
                          0  & \mbox{for $a = k+1, ..., n$}
                          \end{array}
                 \right.
\ee
For the corresponding Hamiltonian and the saddle-point equation
(in the limit $n \to 0$) one gets:

\be
\lb{abbcc}
H_{k} = \fr{1}{\a} k \f^{\a} -
\2 \b h_{0}^{2} k^{2} \f^{2}
\ee
\be
\lb{abbd}
\f^{\a-1} - \b h_{0}^{2} k \f = 0
\ee
The solution on this equation and the corresponding energy are:
\be
\lb{abbf}
\f = (\b k h_{0}^{2})^{\fr{1}{\a-2}}
\ee
\be
\lb{abbg}
H_{k} = - \fr{1}{\b} \fr{\a - 2}{2\a}
(\b k)^{\fr{2(\a - 1)}{\a - 2}} h_{0}^{\fr{2\a}{\a - 2}}
\ee
(Note, that although one can try with more "RSB" steps in the replica
vector $\f_{a}$ it can be easily proved that there exists only one
type of the non-trivial solution given by the Ansatz (\ref{abbe}) ).
Then, in terms of the procedure described above for the free
energy we have:

\be
\lb{abbh}
F(h_{0}) = -\fr{1}{\b} \sum_{k=1}^{\infty} \fr{(-1)^{k-1}}{k}
\exp\{ \fr{\a - 2}{2\a}
(\b k)^{\fr{2(\a - 1)}{\a - 2}} h_{0}^{\fr{2\a}{\a - 2}} \}
\ee
The summation of this series can be performed in terms of the
integral representation eq.(\ref{abj}):

\be
\lb{abbi}
F(h_{0}) = \fr{1}{2i\b} \int_{C} \fr{dz}{z\sin(\pi z)}
\exp\{ \fr{\a - 2}{2\a}
(\b z)^{\fr{2(\a - 1)}{\a - 2}} h_{0}^{\fr{2\a}{\a - 2}} \}
\ee
where the integration goes over the  contour in the complex plane
shown in Fig.1a. Then, again, we move the  contour to the position shown
in Fig.1b and redefine the integration variable:

\be
\lb{abbj}
z \to \fr{1}{\b} h_{0}^{-\fr{\a}{\a - 1}} i x
\ee
In the limit $\b \to \infty$ we have:

\be
\lb{abbk}
\sin(\pi z) \simeq \fr{1}{\b} i \pi h_{0}^{-\fr{\a}{\a - 1}} x
\ee
and

\be
\lb{abbl}
F(h_{0}) = - h_{0}^{\fr{\a}{\a - 1}} \fr{1}{2\pi}
\int_{C_{1}} \fr{dx}{x^{2}}
\exp\{ \fr{\a - 2}{2\a}
(ix)^{\fr{2(\a - 1)}{\a - 2}}  \} \equiv B(\a) \times h_{0}^{\fr{\a}{\a - 1}}
\ee
Thus, we have obtained the correct scaling of the free energy as a function
of $h_{0}$. Note however, that although it is also possible to
calculate the value of the coefficien $B(\a)$ in the integral (\ref{abbl}),
such a calculation would not make much  sense because to obtain the correct
coefficient (which is given by the integral in (\ref{abbb})) one
would need to take into account replica fluctuations which we have
neglected here.

\subsection{The Toy Model}

Let us consider now a slightly less trivial example of a zero-dimensional
system which cannot be solved by elementary algebra. This
system, generally called the "toy model", consists of a single
degree of freedom $\f$ evolving in an energy landscape which is
the sum of a quadratic well and a Brownian potential. The
Hamiltonian is:

\be
\lb{aca}
H = \2\mu\f^{2} + V(\f)
\ee
where $V(\f)$ is the random potential described by the Gaussian
distribution:

\be
\lb{acb}
P[V(\f)] \sim \exp\{-\fr{1}{4g}\int d\f (\fr{dV}{d\f})^{2} \}
\ee
The $V$ distribution is characterized by its first two moments:
\be
\lb{acc}
\ba
\ol{(V(\f) - V(\f'))^{2}} = 2g |\f - \f'| ; \\
\\
\ol{V(\f)} = 0 ;\\
\\
\ol{V(\f)V(\f')} = C - g |\f - \f'|
\ea
\ee
where $C$ is an irrelevant constant
This problem was introduced originally as a toy, zero
dimensional version of the interface in the random field Ising
model \cite{Villain}. It has the virtue of showing explicitely how the most
standard field theoretic methods like perturbation theory,
iteration methods or supersymmetry
 get fooled in this problem, in the low $\mu$,
low temperature limit, by the existence of many metastable states
\cite{Villain,Villain_pert,engel,Villain_iter}.
The main point is that at low enough temperatures the usual perturbation
theory does not work and a qualitatively
reasonable theory must involve the effects of the replica symmetry breaking.
This has been demonstrated  within  the replica
gaussian variational  approximation \cite{mptoy,engel2}.

One quantity which one would like to calculate in
such a system is the value of $\ol{\la\f^{2}\ra}$ in the limit of
the zero temperature. Using simple energy
arguments one can easily estimate  what must be the
scaling dependence of this quantity on the parameters $\mu$ and $g$.
For a given value of $\f$ the loss of energy due to the attractive
potential in the Hamiltonian (\ref{aca}) is $\sim \mu \f^{2}$. Possible
gain of energy due to the random potential according to statistics
(\ref{acb})-(\ref{acc}) can be estimated as $\sim \sqrt{g}\sqrt{\f}$.
Optimizing the total energy $E \sim \mu \f^{2} - \sqrt{g}\sqrt{\f}$
with respect to $\f$ one finds that

\be
\lb{acd}
\ol{\la\f^{2}\ra} =C_2 \  \fr{g^{2/3}}{\mu^{4/3}}
\ee
This result tells that the typical energy minimum of the Hamiltonian
(\ref{aca}) lies at a finite distance from the origin. The
scaling (\ref{acd}), which is obviously right, is not so easy to derive
from some field theoretic methods which could be also used in higher
dimension, and there is no known exact result
for the constant $C_2$ at the moment.

Let us try to calculate the value of $\ol{\la\f^{2}\ra}$ in the zero
temperature limit using the method considered above.
The replicated Hamiltonian of the system (\ref{aca}) is:

\be
\lb{ace}
H_{n} = \2\mu\sum_{a=1}^{n}\f^{2}_{a}
+ \2\b g \sum_{a,b=1}^{n} |\f_{a} - \f_{b}|
\ee
The corresponding saddle-point equations are:

\be
\lb{acf}
\mu \f_{a} + \b g \sum_{b=1}^{n} Sign (\f_{a} - \f_{b}) = 0
\ee

(Note that in this formula, whenever there is some ambiguity, one
should always assume that there is at some intermediate step a short scale
regularization. Therefore, one must interpret for instance
$Sign(0)=0$.)
Let us first look for  non-trivial solutions of the eqs.(\ref{acf}).
It can be easily proven that within the "one-step" RSB Ansatz (\ref{abe})
there exist no non-trivial solutions. Let us consider the "two-steps"
Ansatz for the replica vector $\f_{a}$:

\be
\lb{acg}
\f_{a} = \left\{ \begin{array}{ll}
                 \f_{1} & \mbox{for $a = 1, ..., k$}\\
                          \f_{2} & \mbox{for $a = k+1, ..., k+l$}\\
                 \f_{3} & \mbox{for $a = k+l+1, ..., n$}
                          \end{array}
                 \right.
\ee
 From eqs.(\ref{acf}) one finds the following equations for $\f_{1,2,3}$
(in the limit $n \to 0$):

\be
\lb{ach}
\ba
\mu \f_{1} + \b g l Sign (\f_{1} - \f_{2}) -
\b g (k+l) Sign (\f_{1} - \f_{3}) = 0 \\
\\
\mu \f_{1} + \b g k Sign (\f_{2} - \f_{1}) -
\b g (k+l) Sign (\f_{2} - \f_{3}) = 0 \\
\\
\mu \f_{3} + \b g k Sign (\f_{3} - \f_{1}) +
\b g l Sign (\f_{3} - \f_{2}) = 0
\ea
\ee
The solution of these equations is:

\be
\lb{aci}
\f_{1} = - \fr{g}{\mu}\b k \; ; \; \; \; \; \;
\f_{2} = + \fr{g}{\mu}\b l \; ; \; \; \; \; \;
\f_{3} =  \fr{g}{\mu}\b (l-k)
\ee
and  the corresponding energy is (in the limit $n \to 0$):

\be
\lb{acj}
E_{kl} = -\fr{\b^{2} g^{2}}{2\mu} kl(k+l)
\ee

It can be proven that there exist no other solutions
of the saddle-point equations (\ref{acf}) with a number of  RSB
steps larger than two.

Therefore, (after taking the limit $n \to 0$) for the "RSB" part of the free
energy we get the following series (see eq.(\ref{abu})):

\be
\lb{ack}
\ba
F_{rsb} = -\fr{1}{\b n} \sum_{k+l=1}^{n} \fr{n!}{k!l!(n-k-l)!}
\exp(-\b E_{kl}) \to \\
\\
-\fr{1}{\b} \sum_{k+l=1}^{\infty} \fr{(-1)^{k+l-1}}{k+l}
\fr{(k+l)!}{k!l!}
\exp\{ \lm kl(k+l) \}
\ea
\ee
where

\be
\lb{acl}
\lm = \fr{\b^{3} g^{2}}{2\mu} \to \infty
\ee

We again carry the summation of the asymptotic series (\ref{ack}) with
 the integral method mentioned in section 2.2:

\be
\lb{acm}
F_{rsb} = \fr{1}{\b(2i)^{2}} \int \int_{C}
\fr{dz_{1} dz_{2}}{(z_{1}+z_{2})\sin(\pi z_{1})\sin(\pi z_{2})}
\fr{\G(z_{1}+z_{2}+1)}{\G(z_{1}+1)\G(z_{2}+1)}
\exp\{ \lm z_{1}z_{2}(z_{1}+z_{2})\}
\ee
where the integrations over $z_{1,2}$ both go around the  contour in the
complex
plain shown in Fig.1a.

Shifting the  contour of integration to the position shown in Fig1.b,
and redefining the integration variables: $z_{1,2} \to \lm^{-1/3} i x_{1,2}$
in the limit $\b \to \infty$ ($\lm^{-1/3} \to 0$) one gets:
\be
\lb{aco}
F_{rsb} = \fr{1}{\b}\fr{\lm^{1/3}}{2\pi^{2}}
\{\int\int_{0}^{+\infty} dx_{1} dx_{2} [
\fr{\sin(x_{1}x_{2}(x_{1}+x_{2}))}{x_{1}x_{2}(x_{1}+x_{2})} +
\fr{\sin(x_{1}x_{2}(x_{1}-x_{2}))}{x_{1}x_{2}(x_{1}-x_{2})}] \}
\ee
Substituting the value of $\lm = \b^{3} g^{2}/2\mu$
we finally get the result for the zero-temperature free energy:

\be
\lb{acp}
 F_{rsb} = \fr{g^{2/3}}{\mu^{1/3}}
\fr{\sqrt{3} \G(1/6)}{4 \pi^{3/2}}
\ee

To this piece we must now add the replica symmetric contribution.
The saddle point equations have the trivial solution: $\f_{a} = 0$
with the corresponding energy $E_{0} \equiv H_{n}[\f_{a}=0] = 0$. As
we want to get a quantitative result for the constant $C_2$, we
must also include the contribution from the replica
fluctuations around this saddle point. This cannot be done
just at the level of integrating the quadratic fluctuations.
We shall rather make the following (strong) assumption, namely that
this whole `RS' part of the free energy, including the
replica fluctuations, is given by the Gaussian replica variationnal
method \cite{MP,mptoy,engel2}. We do not have a very convincing
argument to support this hypothesis; we just point out that
this gaussian variationnal method involves the Gaussian integration over
replica
fields which in a  sense is "symmetric" with respect to the point
$\f_{a} = 0$. In the end the hypothesis is best supported by the good result
one
gets for $C_2$. We denote the gausian variational contribution by $F_{rv}$,
and our conjecture is that $F=F_{rv}+F_{rsb}$.

According to eq.(\ref{aca}):

\be
\lb{acr}
\ol{\la\f^{2}\ra} = 2 \fr{\partial F}{\partial \mu} =
\ol{\la\f^{2}\ra}_{rv} - \fr{g^{2/3}}{\mu^{4/3}}
\fr{\G(1/6)}{2\sqrt{3} \pi^{3/2}}
\ee
Using the result of \cite{mptoy} for the value of $\ol{\la\f^{2}\ra}_{rv}$
we finally get:

\be
\lb{acs}
\ol{\la\f^{2}\ra}  =
\fr{g^{2/3}}{\mu^{4/3}} \((\fr{3}{(4\pi)^{1/3}} -
 \fr{\G(1/6)}{2\sqrt{3} \pi^{3/2}} \)) \simeq 1.00181 \fr{g^{2/3}}{\mu^{4/3}}
\ee

We have compared this result with some numerical simulations
of the problem.
The scaling in $\mu$ and $g$ is obviously correct, the only point to
check is the prefactor $C_2$. Choosing for instance
 the values of the parameters $\mu = 1$ and
$g = 2\sqrt{\pi}$ (when the replica variational method gives
$\ol{\la\f^{2}\ra}_{rv} = 3$ )
we obtain from (\ref{acs}) the analytical prediction: $\ol{\la\f^{2}\ra} \simeq
2.3291$.
The numerical simulation was done at zero temperature, with these
same values of
 $\mu$ and
$g$. The $\f$ interval $[-8,8]$ is discretized in $2 N$ points,
on which one generates a random potential as in (\ref{aca}). The exhaustive
scan
gives the minimum, from which one computes $\ol{\la\f^{2}\ra}$.
We average over $100000$ samples. The number of points
$2 N$ ranged from $2^8$ to $2^{16}$, in this regime there is no
systematic
$N$ dependance.  There is
no systematic error due to the finite width of the interval since we have
checked that, within our statistics, there is no sample for which the
minimum is found with $|\f| \ge 7$. The result of the simulation
is $ \ol{\la\f^{2}\ra} \simeq 2.45 \pm .02$.
 The value predicted by our replica saddle point summation is rather close,
although there is a clear small discrepancy.

\vspace{5mm}

{\it The result for $\ol{\la\f^{4}\ra}$}

\vspace{5mm}

To be sure that this relatively
good agreement of our prediction for $\ol{\la\f^{2}\ra}$
with the numerical result is not just a coincidence we have performed similar
calculations
for the  next order correlator $\ol{\la\f^{4}\ra}$. The computations, which are
similar
to the ones we have just presented but more cumbersome, are given in the
appendix.
The result is:

\be
\lb{acy}
\ol{\la\f^{4}\ra} = \ol{\la\f^{4}\ra}_{rv} + \ol{\la\f^{4}\ra}_{rsb} =
\fr{g^{4/3}}{\mu^{8/3}} \((\fr{27}{(4\pi)^{2/3}}
- \fr{17\sqrt{3} [\sin(\pi/12) + \cos(\pi/12)]}{
3\sqrt{\pi} \G(1/6) \sin(\pi/6)}\))
\ee
For the values of the parameters $\mu = 1$ and
$g = 2\sqrt{\pi}$ (when the replica variational method gives
$\ol{\la\f^{4}\ra}_{rv} = 27$ )
we obtain: $\ol{\la\f^{4}\ra} \simeq 16.25$. The numerical result is
obtained with the same procedure as above and gives
$\ol{\la\f^{4}\ra} \simeq 17.05 \pm .2$. Again
these number are close but there is a significative difference.

\vspace{5mm}

Clearly, the vector type of rsb that we have been using on all these
zero dimensional problems is somewhat strange, and we cannot assert
that we control all of its aspects (in particular the fact that the
replica fluctuations around the rs saddle point are summed
by the gaussian variational method is still unclear). However in all
these cases, and in particular in the non-trivial case of the
toy model, we have obtained good results using this simple
receipe. Therefore we now turn to its application to more
elaborate problems, starting with systems in one dimension.

\sectio{ Directed Polymers in Random Media}

The problem of a directed polymer in a random medium is an important
problem which has been much studied recently \cite{DPRM}. Although
the situation in 1+1 dimension, with a delta correlated potential,
is relatively well understood, there are still a lot of uncertainties
about more complicated cases.

We shall consider a one dimensional case with long range correlations
of the potential. It is described by a  one-dimensional scalar field system
with
 the following Hamiltonian:

\be
\lb{baa}
H = \int_{0}^{L} dx [\2(\fr{d\f(x)}{dx})^{2} + V(x,\f)]
\ee
where the random potentials $V(x,\f)$ are described by the Gaussian
distribution with {\it non-local} correlations with respect to the
fields $\f$:

\be
\lb{bab}
\ol{V(x,\f)V(x',\f')} = \d(x-x')[const - g (\f - \f')^{2\a}]
\ee
where $0 < \a < 2$.

This problem naturally arises, with $\a=1/2$, when one considers an
interface in the two dimensional random field Ising model at low temperatures:
then the field $\f$ just describes the lateral fluctuations in
the interface, in a solid on solid approximation.

One first basic question that we would like to answer concerns
the scaling behaviour of the lateral fluctuations.
Let the value of the field $\f(x)$ be sticked
to zero at the origine: $\f(x=0) \equiv 0$. Then one would like
to know how the average value of the field at $x=L$,
$\ol{\la\f(L)^{2}\ra}$, scales with $L$:

\be
\lb{bac}
\ol{\la\f(L)^{2}\ra} \equiv
\ol{\left( Z^{-1} \int d\f_{0} \f_{0}^{2}
\int_{\f(0)=0}^{\f(L)=\f_{0}} D\f(x) \exp(-\b H[\f(x),V]) \right) } \sim
L^{2 \zeta}
\ee
where the partition function $Z$ (for a given realization of the random
potential) is given by the integration over all the trajectories $\f(x)$
with only one boundary condition $\f(x=0) = 0$. The `wandering exponent'
$\zeta$
has been computed in the case of local correlations of the random potential,
it is then equal to $2/3$ \cite{DP1+1}. In the case of non local
correlations such as (\ref{bab}), it is believed that this exponent should
be equal to $3/2(2 - \alpha)$ at small enough $\alpha$. This is the
result that is obtained from the gaussian variational Ansatz \cite{MP}, and
it can also be derived from a mapping to the Burgers (or
the KPZ) equation and
a study of this equation through a dynamical renormalization
group procedure \cite{KPZ}.

A simple derivation of this scaling can be obtained by an energy
balance argument a la Imry Ma \cite{ImryMa}.
Let the value of the field be equal
to $\f_{0}$ at $x = L$. Then the loss of the energy due to the
gradient term in the Hamiltonian (\ref{baa}) can be estimated as
$E_{g} \sim \f_{0}^{2}/L$. The gain of energy due to the
random potential term, according to eq.(\ref{bab}), can be estimated
as $E_{V} \sim - \sqrt{L} \sqrt{g} \f_{0}^{\a}$. Optimizing $E_{g}$ and
$E_{V}$ with respect to $\f_{0}$ one finds:

\be
\lb{bad}
\f_{0} \sim L^{\fr{3}{2(2-\a)}} \; g^{\fr{1}{2(2-\a)}}
\ee

In this section we will demonstrate how this result can be obtained
in the zero-temperature limit in terms of the proposed replica saddle point
method. The replicated Hamiltonian is:

\be
\lb{bae}
H_{n} = \int_{0}^{L} dx \left[ \2\sum_{a=1}^{n} (\fr{d\f_{a}(x)}{dx})^{2} +
\2 \b g \sum_{a,b=1}^{n} (\f_{a}(x) - \f_{b}(x))^{2\a} \right]
\ee
Strictly speaking, the systematic way of solving this problem
following our general method is the following: one must find $n$ saddle point
trajectories $\f_{a}(x)$ for fixed $n$ boundary conditions $\f_{a}(L)$,
then one has to derive the corresponding energy
$\tl{H_{n}}[\f_{a}(L)]$, and finally one has to find the saddle point
solutions with respect to the values of $\f_{a}(L)$.

Here we shall follow a much  simpler strategy. Since it is obvious that
there always exists the trivial solution $\f(x) \equiv 0$,
we will suppose that the correct scaling can be obtained simply by
taking into account one  non-trivial saddle-point trajectory. In other words,
from the very begining we are going to look for the saddle point
solutions within the following Ansatz:

\be
\lb{baf}
\f_{a}(x) = \left\{ \begin{array}{ll}
                 \f(x) & \mbox{for $a = 1, ..., k$}\\
                          0     & \mbox{for $a = k+1, ..., n$}
                          \end{array}
                 \right.
\ee
Comparing this Ansatz to the zero dimensional exercises of the previous
section, we see that it should amount to assuming that the lowest
lying configuration dominates. This is certainly true since one knows
\cite{parisi91,mezard91}
that the metastable states have an excitation energy which scales
as $L^\omega$, with $\omega=2 \zeta-1$.
Substituting this Ansatz into the replica Hamiltonian (\ref{bae})
in the limit $n \to 0$ one gets:

\be
\lb{bag}
H_{k} = k \int_{0}^{L} dx \left[ \2 (\fr{d\f(x)}{dx})^{2} -
\b k g \f^{2\a}(x) \right]
\ee
As usual (see the previous section) the free energy is defined by the
series:

\be
\lb{bah}
F(L) \sim -\fr{1}{\b} \sum_{k=1}^{\infty} \fr{(-1)^{k-1}}{k}
\exp(-\b H_{k})
\ee
where the value $H_{k}$ is defined by the corresponding saddle-point solution
for $\f(x)$.

The saddle point trajectory is defined by the following
differential equation:

\be
\lb{bai}
\fr{d^{2}\f}{dx^{2}} = - 2 \a \b k g \f^{2\a - 1}
\ee
with the boundary conditions: $\f(0) = 0$ and $\f(L) = \f_{0}$.
This equation can be easily integrated:

\be
\lb{bak}
\int_{0}^{\f(x)} \fr{d\f}{\sqrt{\lm - \f^{2\a}}} = x \sqrt{2 \b k g}
\ee
where the integration constant $\lm$ is defined by the boundary
condition:

\be
\lb{bal}
\int_{0}^{\f_{0}} \fr{d\f}{\sqrt{\lm - \f^{2\a}}} = L \sqrt{2 \b k g}
\ee
Substituting this solution into the Hamiltonian (\ref{bag}), we obtain
 after some simple algebra:

\be
\lb{bam}
H_{k} = k \left[ -\b k g \lm L + \sqrt{2\b k g}
\int_{0}^{\f_{0}} d\f \sqrt{\lm - \f} \right]
\ee
Taking derivative of $H_{k}$ with respect to $\f_{0}$
(and taking into account the constrain (\ref{bal})) one finds
the following saddle-point solution:

\be
\lb{ban}
\f_{0} \sim L^{\fr{1}{1-\a}} \; (\b k g)^{\fr{1}{2(1-\a)}}
\ee
and $\lm = \f_{0}^{2\a}$. Its energy (\ref{bam}) is:

\be
\lb{bao}
H_{k} = - \fr{(const)}{\b} (\b k)^{\fr{2-\a}{1-\a}} \; L^{\fr{1+\a}{1-\a}}
\; g^{\fr{1}{1-\a}}
\ee
Now we proceed as before, introducing an integral representation of the series
(\ref{bah})
and a rescaling
of the integration  variable by a factor $
{1 \ov \b}  L^{- \fr{1+\a}{2-\a}}  g^{- \fr{1}{2-\a}}
$.
Then we  get the scaling of the
free energy:

\be
\lb{bas}
F(L) \sim L^{\fr{1+\a}{2-\a}}  g^{\fr{1}{2-\a}}
\ee
from which we obtain the scaling of $\f_{0}$ as a function of $L$:

\be
\lb{bat}
\f_{0}(L) \sim L^{\fr{3}{2(2-\a)}}  g^{\fr{1}{2(2-\a)}}
\ee
which  coincides with the result (\ref{bad}) given by the naive energy
arguments, as well as by more elaborate calculations.

Although the example demonstrated in this section provides no new
results we hope that the proposed method could turn out to be also useful
when applied for  directed polymers with smaller $\alpha$, or in larger
dimension.

\sectio{Random Field Ising Model in $D$ dimensions}

Since the topic of the random field Ising model covers an enormous amount
of litterature (see e.g \cite{rfimrev}), it would be rather difficult to give
any brief introductory review. Here, however, we are mainly concerned
with how the  method we have proposed before works in various situations.
Therefore,  we will concentrate only onto one particular aspect
of the problem.

It is well known that the main problem in the studies of the
low temperature phase in the random field Ising model is that
one has to perform the summation over numerous local minima states,
which seems to be impossible to do within the framework of the usual
perturbation theory \cite{rfimrev}. It has been proposed
recently that, because of these
local minima states a special "intermediate" (separating paramagnetic
and ferromagnetic phase) spin-glass-like thermodynamic state could set in
around
the critical point, and moreover, this state is characterized by
a replica symmetry breaking in the corresponding correlation
functions \cite{mezyoung}. At low temperature, and when the width
of the distribution of the random field is not too
small, the same phenomenon must be present. It was proposed
long ago \cite{parHouches}, and elaborated later on
in \cite{dotpar}, that the metastable states in this regime
should be characterized by some ``instanton in replica space".
Our method provides one more step in the elaboration
of this idea.

We consider the random field Ising model in terms of the usual
Ginzburg-Landau Hamiltonian in $D$ dimensions:

\be
\lb{caa}
H = \int d^{D}x \left[ \2 (\n\f)^{2} + \2\t\f^{2} + \4 g \f^{4}
- h(x)\f\right]
\ee
where the random fields $h(x)$ are described by the $\d$-correlated
Gaussian distribution:

\be
\lb{cab}
P[h(x)] = \prod_{x} \left[\fr{1}{\sqrt{2\pi h_{0}^{2}}}
\exp\((-\fr{h^{2}(x)}{2 h_{0}^{2}} \)) \right]
\ee
The corresponding replica Hamiltonian is:

\be
\lb{cad}
H_{n} = \int d^{D}x \left[ \2 \sum_{a=1}^{n} (\n\f_{a})^{2} +
\2 \t \sum_{a=1}^{n} \f_{a}^{2} + \4 g \sum_{a=1}^{n} \f_{a}^{4}
- \2 h_{0}^{2} \sum_{a,b=1}^{n} \f_{a} \f_{b} \right]
\ee
According to the procedure developed in previous sections we are going to
look for the most simple non-trivial saddle-point solutions at the background
of the trivial one, $\f_{a}(x) \equiv 0$. In terms of the Ansatz:

\be
\lb{cae}
\f_{a}(x) = \left\{ \begin{array}{ll}
                 \f(x) & \mbox{for $a = 1, ..., k$}\\
                          0     & \mbox{for $a = k+1, ..., n$}
                          \end{array}
                 \right.
\ee
The replica Hamiltonian (\ref{cad})  reads in the $n\to 0$ limit:

\be
\lb{caf}
H_{k} = k \int d^{D}x \left[ \2 (\n\f)^{2} -
\2 (h_{0}^{2} k - \t) \f^{2} + \4 g \f^{4} \right]
\ee
Consider for simplicity the situation at $\t = 0$
(Notice that here we work close to the critical temperature.
The use of our saddle point technique allows to
study the system at the tree level, which is supposed to give the leading
singularities close to $T_c$ \cite{youparsou}).
The corresponding
saddle-point equation is:

\be
\lb{cag}
-\D \f - \lm \f + g \f^{3} = 0
\ee
where $\lm = h_{0}^{2} k$. As usual,  the free energy
is given by the series:

\be
\lb{cah}
F(h_{0}) \sim - \sum_{k=1}^{\infty} \fr{(-1)^{k-1}}{k} \exp(- H_{k})
\ee
where the value of $H_{k}$ is defined by the corresponding saddle-point
solution
of the eq.(\ref{cag}).

At this stage we see that the situation is getting rather different from the
ones studied in the previous sections. If we would choose the obvious
space-independent solution $\f = (\lm/g)^{1/2}$, we would find that
the value of $H_{k}$ is proportional to the volume $V$ of the system:
$H_{k} = -\4 k (\lm^{2}/g) V = -\fr{1}{4 g} k^{3} h_{0}^{4} V$.
Then, the summation of the series (\ref{cah}) would immediately yield
a free energy  proportional to $V^{1/3}$ and not to $V$.
Therefore this solution, as well as any other solution with an energy
$H_{k}$  proportional to the volume of the system, is irrelevant
for the bulk properties.

Thus, we have to look for  {\it localized} solutions: the ones
which are local in space (breaking translation invariance) and which have
{\it finite} energy. Let us assume to start with
that such an "instanton"-type solution exists
(see below), and that for a given $k$ it is characterized by the spatial size
$R(k)$. Then, if we take into account only one-instanton contribution
(or in other words we consider a gas of {\it non-interacting} instantons),
due to the trivial entropy factor $V/R^{D}$ (this is the number of positions
of the object of the size $R$ in the volume $V$) we  get a
free energy  proportional to the volume:

\be
\lb{cai}
F(h_{0}) \sim - \sum_{k=1}^{\infty} \fr{(-1)^{k-1}}{k} \fr{V}{R^{D}}
\exp(- H_{k})
\ee
where $H_{k}$ must be finite (volume independent).

It is easy to understand that the equation (\ref{cag}) indeed has
 localized solutions. Let us assume that
the value of the field $\f(x)$
is such that $\lm \f^{2} \gg g \f^{4}$. Then in a first approximation
the saddle-point equation (\ref{cag}) is  linear:

\be
\lb{caj}
\D \f + \lm \f = 0
\ee
The simplest possible spherically-symmetric solutions of this equation in
$D$ dimensions are the well known Bessel-type functions. In particular
there exist oscillating solutions which have a finite value
$\f(r=0) \equiv \f_{0}$ at the origin and which decay to zero at
$r \to \infty$ (like $\sim r^{-(D-1)/2} \sin r$). For example,
in dimension $D = 3$ this solution is simply:

\be
\lb{cak}
\f(r) = \f_{0} \fr{\sin(r\sqrt{\lm})}{r\sqrt{\lm}}
\ee

In dimensions $D$ these solutions have a finite spatial scale:

\be
\lb{cakk}
R(k) = \lm^{-1/2} = (h_{0}^{2} k)^{-1/2}
\ee
and finite energy:

\be
\lb{cal}
H_{k} = - (const) \; k \f_{0}^{2} \lm^{-\fr{D-2}{2}}
\ee
At the level of the equation (\ref{caj}) itself, the value of $\f_{0}$ remains
arbitrary (the equation is linear). On the other hand, from the point of
view of the energy this is not an extremum since the energy explicitely
depends on the value of $\f_{0}$ (this is the saddle-point solution for
the fixed boundary condition $\f(r=0) = \f_{0}$). If we would let the value
of $\f_{0}$  be free in the absence of the non-linear term $g\f^{4}$
it would, of course, fall down to infinity. However, if we take into account
the term $g\f^{4}$ in the "exact" Hamiltonian" (\ref{caf}) it is natural
to expect that $\f_{0}$ will stabilize around the saddle-point value

\be
\lb{cam}
\f^{2}_{0} = \fr{\lm}{g}
\ee
The above qualitative arguments can be easily verified for the model
double-well potential: $\tl{U}(\f) = -\2\f^{2}$ for
$|\f| \leq \sqrt{\lm/g}$ and $\tl{U}(\f) = +\infty$ for
$|\f| > \sqrt{\lm/g}$, taken instead of the "real" one:
$U(\f) = -\2\f^{2} + \4 g\f^{4}$. In this case for any
$|\f_{0}| \leq \sqrt{\lm/g}$ there exists the exact Bessel-like saddle-point
solution with finite energy (\ref{cal}), and real extremum of the Hamiltonian
would be achieved at $\f_{0} = \pm \sqrt{\lm/g}$.

Let us calculate the contribution of such solutions to the free energy.
Substituting into the series (\ref{cai}) the energy of the solution
(\ref{cal}), the estimate for the value of $\f_{0}$ (\ref{cam})
and the characteristic size of the solution (\ref{cakk}),
together with $\lm = h_{0}^{2} k$ we get:

\be
\lb{can}
F(h_{0}) \sim - V \sum_{k=1}^{\infty} \fr{(-1)^{k-1}}{k}
(h_{0}^{2} k)^{\fr{D}{2}}
\exp\left[ \fr{(const)}{g h_{0}^{2}} (h_{0}^{2} k)^{\fr{6-D}{2}} \right]
\ee
We see that the series is getting strongly divergent only
 at dimensions $D < 6$.
This is the only regime where
 the considered saddle-point
solutions provide a relevant contribution.

Now, following the scheme developped in the previous sections,
we turn to the integral representation and rescale the integration
variable by a factor
$
(g h_{0}^{2})^{\fr{2}{6-D}} h_0^{-2}
$, which gives a free energy with the following
scaling in the limit $g h_{0}^{2} \ll 1$:

\be
\lb{car}
\fr{F(h_{0})}{V} \sim \fr{1}{g} (g h_{0}^{2})^{\fr{4}{6-D}}
\ee
Besides, using the same scaling
$ k \sim (g h_{0}^{2})^{\fr{2}{6-D}} h_0^{-2} $
for the characteristic spatial scale of
the saddle-point solutions (\ref{cakk}), which could be interpreted as a kind
of  disorder induced {\it finite} correlation length
(near $T = \T$, as we shall see in more details below), we obtain:

\be
\lb{cas}
R_{c}(h_{0}) \sim (g h_{0}^{2})^{-\fr{1}{6-D}}
\ee
In the same way one gets the estimate for the value of the "disorder
parameter" $\ol{\f^{2}} \sim \f_{0}^{2} \simeq  \fr{1}{g}(h_{0}^{2} k)$:

\be
\lb{cass}
\f_{0}^{2} \sim \fr{1}{g} (g h_{0}^{2})^{\fr{2}{6-D}}
\ee
Finally, one can easily obtain the estimate for the value of the
temperature interval $\t_{c}$ around $\T$ where all the above qualitative
calculations
make sense. Formally the derivation of the
saddle-point solutions has been done for $\t = 0$. Actually, according to
the replica Hamiltonian (\ref{caf}) the calculations should remain
 correct until $|\t| \ll h_{0}^{2} k$. Using the above scale estimate
for $k$ one finds the upper bound for the value of $\t$:

\be
\lb{cat}
|\t| \ll \t_{c} \sim (g h_{0}^{2})^{\fr{2}{6-D}}
\ee
This value of $\t_{c}$ can be interpreted as the estimate for the temperature
interval around $\T$ where the supposed disorder dominated (spin-glass type)
phase sets in.

Of course, the procedure proposed in this section is still incomplete.
 In a selfconsistent approach one should  study
 the effects produced by the interactions between these instanton
solutions, not talking about the effects of the critical fluctuations.
At the present stage we are not able to say anything about the
ferromagnetic phase transition itself and in particular about the behaviour
of the corresponding ferromagnetic order parameter.

Nevertheless, we shall now show that these simple replica instanton estimates
are quite reasonable and can in fact be recovered in terms of (completely
independent) simple scaling arguments.
Indeed, let us come back to the original random field Hamiltonian
(\ref{caa}). Configurations of the field $\f(x)$ which correspond to
local minima satisfy the saddle-point equation:

\be
\lb{cbaa}
-\D\f(x) + \t\f(x) + g\f^{3}(x) = h(x)
\ee
Let us estimate at which spatial and temperature scales the random fields
give a dominant contribution.
We consider a large region $\Omega_L$ of  linear size $L \gg 1$.
The spatially averaged value of the random field in this region is:

\be
\lb{cbab}
h(\Omega_L) \equiv \fr{1}{L^{D}} \int_{x\in\Omega_{L}} d^{D}x h(x)
\ee
Correspondingly, the typical average value of the random field in this region
of  size $L$ is:

\be
\lb{cbac}
h_{L} \equiv \[[\ol{ h^{2}(\Omega_{L})}\]]^{1/2} = h_{0} L^{-D/2}
\ee
Then the estimate for the typical value of the order parameter field $\f_L$
in this region can be obtained from the saddle-point equation:

\be
\lb{cbad}
\t \f_{L} + g \f^{3}_{L} = h_{L}
\ee
Then, as long as:

\be
\lb{cbae}
\t\f_{L} \ll g \f^{3}_{L}
\ee
the typical value of $\f_{L}$ inside such clusters
is dominated by the random field:

\be
\lb{cbb}
\f_{L} \sim (h_{L}/g)^{1/3} \sim
(\fr{h_{0}}{g})^{1/3} L^{-D/6}
\ee

Now let us estimate up to which characteristic size of the cluster the
external fields can dominate. According to (\ref{cbae}) and (\ref{cbb})
one gets:

\be
\lb{dddk}
L << \frac{(g h_{0}^{2})^{1/D}}{\t^{3/D}}
\ee

On the other hand, the estimation of the order parameter in terms of the
equilibrium equation (\ref{cbad}) can be correct only on length scales
much larger than the size of the fluctuation region which is equal to the
correlation length (of the pure system) $\R \sim \t^{-\nu}$.
Thus, one has the lower bound for $L$:

\be
\lb{dddl}
L >> \t^{-\nu}
\ee

Therefore, the region of parameters  where the external fields
dominate is:

\be
\lb{dddm}
\t^{-\nu}<< \frac{(g h_{0}^{2})^{1/D}}{\t^{3/D}}
\ee
or

\be
\lb{dddn}
\t^{3 - \nu D} << g h_{0}^{2}
\ee
Such a region of temperatures near $\T$ exists only if:

\be
\lb{dddo}
\nu D < 3
\ee
In this case the temperature interval near $\T$ in which the order parameter
configurations are mainly defined by the random fields is:

\be
\lb{dddp}
\t_{c}(h_{0}) \sim (g h_{0}^{2})^{\frac{1}{3-\nu D}}
\ee

In the mean field theory (which correctly describes the phase transition
in the pure system for $D > 4$) $\nu = 1/2$. Thus, according to the condition
(\ref{dddo}) the above non-trivial temperature interval $\t_{c}$ exists
only at dimensions $D < 6$. Substituting $\nu = 1/2$ into (\ref{dddp}) we
get:

\be
\lb{dddpp}
\t_{c}(h_{0}) \sim (g h_{0}^{2})^{\frac{2}{6-D}}
\ee

Then, the random field defined spatial scale can be estimated from
(\ref{dddk}):

\be
\lb{dddq}
L_{c}(h_{0}) \sim (g h_{0}^{2})^{-\fr{1}{6-D}}
\ee

Correspondingly, the typical value of the order parameter field at scales
$L_{c}(h_{0})$ is obtained from the eq.(\ref{cbb}):

\be
\lb{dddr}
\f_{L_{c}}^{2} \sim \fr{1}{g} (g h_{0}^{2})^{\fr{2}{6-D}}
\ee

The energy density is estimated as $\fr{E}{V} \sim \f_{L_{c}} h_{L_{c}}$.
Taking into account (\ref{cbac}) and (\ref{dddr}) we find:

\be
\lb{ddds}
\fr{E}{V} \sim \fr{1}{g} (g h_{0}^{2})^{\fr{4}{6-D}}
\ee

We see that we get  through these simple arguments a region around
$\T$ where the disorder induces a finite correlation length. Furthermore
the estimates for $\fr{E}{V}, \L_{c}, \f_{L_{c}}$ and $\t_{c}$
perfectly  coincide with the results
obtained in terms of our previous replica saddle-point method,
eqs.(\ref{car})-(\ref{cat}). Both approaches clearly hold only in a regime
where
critical fluctuations can be neglected.

\sectio{Conclusions}
We have proposed a  method to analyse random systems by summing
up various saddle point contributions in the replicated Hamiltonian.
We think that it may open a new route in this type of study. In
 particular, the application to finite dimensional systems,
which we started here with the directed polymer on one hand, and with
the random field Ising model on the other hand, looks quite interesting.
Indeed we have seen on this last case how this method allows to
take into account instanton contributions which are usually out of reach
of most analytic methods in these systems. Such instanton contributions
have been argued to be important for a long time (\cite {parHouches,dotpar}).
We think we can get them under control with the present approach.

Clearly our method is still not totally understood in all details.
We have pointed out that it involves one single basic rule, stating the
way one has to order the various saddle points in replica space. Within
this hypothesis it gives reasonable results in all the cases we have
checked so far, but of course more studies are needed to justify
this hypothesis.

\section{Appendix: Computation of the fourth moment in the toy model}
Using the saddle point solution (\ref{aci}) we have:

\be
\lb{act}
\ba
\ol{\la\f^{4}\ra}_{rsb} =
\sum_{k+l=1}^{n} \fr{n!}{k!l!(n-k-l)!}
\left[ k \f_{1}^{4} + l \f_{2}^{4} +(n-k-l) \f_{0}^{4}\right]
\exp\{-\b E_{kl} \} \to \\
\\
 (\fr{\b g}{\mu})^{4}
\sum_{k+l=1}^{\infty} \fr{(-1)^{k+l-1}}{k+l} \fr{(k+l)!}{k!l!}
k l (k+l) (3 k^{2} + 3 l^{2} - 5 kl) \exp\{ \lm kl(k+l) \}
\ea
\ee
Proceeding similarly to the calculations of the free energy $F_{rsb}$
(\ref{acm})-(\ref{aco}) we get:

\be
\lb{acu}
\ba
\ol{\la\f^{4}\ra}_{rsb} =
(\fr{\b g}{\mu})^{4} \fr{\partial}{\partial \lm} \{
-\fr{1}{(2i)^{2}} \int\int_{C}
\fr{dz_{1} dz_{2}}{(z_{1}+z_{2})\sin(\pi z_{1})\sin(\pi z_{2})}
\fr{\G(z_{1}+z_{2}+1)}{\G(z_{1}+1)\G(z_{2}+1)}
(3z_{1}^{2} + 3z_{2}^{2} - 5z_{1}z_{2}) \times \\
\\
\exp[ \lm z_{1}z_{2}(z_{1}+z_{2})] \}
\ea
\ee
Shifting  contour to the position in Fig.1b and redefining
$z_{1,2} \to \lm^{-1/3} i x_{1,2}$
in the limit $\b \to \infty$ ($\lm^{-1/3} \to 0$) we find:

\be
\lb{acv}
\ba
\ol{\la\f^{4}\ra}_{rsb} =
(\fr{\b g}{\mu})^{4} \fr{\partial}{\partial \lm}\{
-\fr{\lm^{-1/3}}{2\pi^{2}}
\int\int_{C_{1}}
\fr{dx_{1} dx_{2}}{(x_{1}+x_{2})x_{1}x_{2}}
(3x_{1}^{2} + x_{2}^{2} - 5x_{1}x_{2}) \times \\
\\
\exp[ -i \lm x_{1}x_{2}(x_{1}+x_{2})] \}
\ea
\ee
Taking into account the contribution from the pole at $x_{1,2} = 0$
after somewhat painful algebra we finally obtain the following result:

\be
\lb{acw}
\ol{\la\f^{4}\ra}_{rsb} =
-\fr{g^{4/3}}{\mu^{8/3}} \fr{17\sqrt{3} [\sin(\pi/12) + \cos(\pi/12)]}{
3\sqrt{\pi} \G(1/6) \sin(\pi/6)}
\ee
Taking into account the contribution from the replica
fluctuations \cite{mptoy}:

\be
\lb{acx}
\ol{\la\f^{4}\ra}_{rv} =
\fr{g^{4/3}}{\mu^{8/3}} \fr{27}{(4\pi)^{2/3}}
\ee
for the fourth order correlator we get the final result:

\be
\lb{acy2}
\ol{\la\f^{4}\ra} = \ol{\la\f^{4}\ra}_{rv} + \ol{\la\f^{4}\ra}_{rsb} =
\fr{g^{4/3}}{\mu^{8/3}} \fr{27}{(4\pi)^{2/3}}
-\fr{g^{4/3}}{\mu^{8/3}} \fr{17\sqrt{3} [\sin(\pi/12) + \cos(\pi/12)]}{
3\sqrt{\pi} \G(1/6) \sin(\pi/6)}
\ee

{\bf Figure captions}

{\bf Fig.1} The contours of integration in the complex plane used for summing
the series.
a) The original contour.  b) The deformed contour.

\epsfxsize=16 cm  \epsfbox{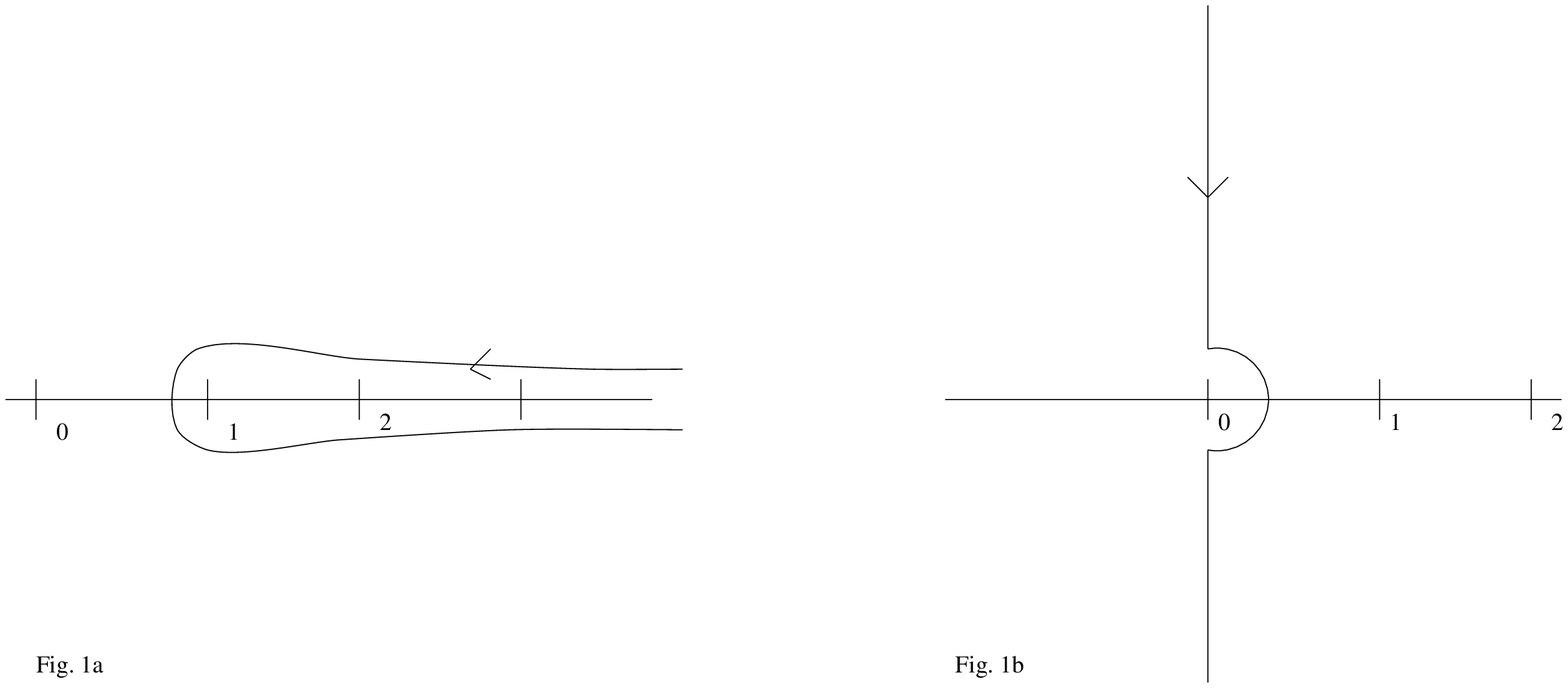}
\end{document}